\documentclass[10pt,conference]{IEEEtran}
\IEEEoverridecommandlockouts

\usepackage{cite}
\usepackage{amsmath,amssymb,amsfonts}
\usepackage[ruled,vlined]{algorithm2e}
\usepackage{quantikz}
\usepackage{adjustbox}

\usepackage{graphicx}
\usepackage{textcomp}
\usepackage{hyperref}
\usepackage{xcolor}
\def\BibTeX{{\rm B\kern-.05em{\sc i\kern-.025em b}\kern-.08em
    T\kern-.1667em\lower.7ex\hbox{E}\kern-.125emX}}
\begin{document}

\title{Q-DICE: Quantum Distributed Interconnect Compiler and Emulator\\
}

\author{\IEEEauthorblockN{Michael Silver, Zachary Vernec, Hans-Arno Jacobsen}
\IEEEauthorblockA{\textit{Edward S. Rogers Sr. Department of Electrical \& Computer Engineering} \\
\textit{University of Toronto}\\
Toronto, Canada \\
\{m.silver, zachary.vernec\}@mail.utoronto.ca, jacobsen@eecg.toronto.edu}
}

\maketitle

\begin{abstract}
As distributed quantum computing (DQC) offers a leading path towards scalable quantum computation, the ability to benchmark distributed algorithms under realistic conditions becomes critical for system co-design. However, without access to physical systems, researchers lack tools to evaluate distribution protocols. We introduce Q-DICE (Quantum Distributed Interconnect Compiler and Emulator), a hardware-aware emulation environment for benchmarking distributed quantum circuits on classical simulators and on NISQ-era monolithic hardware.\\
This work provides three core contributions: (1) a programmatic scheme to construct distributed QPU backends, utilizing two novel techniques---QPU slicing and stitching---to facilitate distributed circuit mapping, (2) a methodology for modeling nonlocal link noise using physically motivated Kraus operators and stochastic error channels, and (3) a boundary-aware circuit mapping algorithm enforcing distributed QPU topology constraints during transpilation. Together, these components constitute a distribution-aware compiler and noise-modeling engine that faithfully enforces the physical limitations of distributed quantum hardware within existing execution environments.\\
We validate Q-DICE against a multitude of experimentally demonstrated quantum circuits, including a distributed Grover's search on optically linked trapped-ion hardware, achieving a worst-case fidelity deviation of 4\% between simulated and experimental results. These findings demonstrate Q-DICE's capacity to accurately reproduce real distributed quantum system behavior across platforms, streamlining experimentation with distributed quantum algorithms and architectures.
\end{abstract}


\section{Introduction}
The scaling of quantum computers toward the fault-tolerant era is currently hindered by the physical and engineering bottlenecks of monolithic quantum processing units (QPUs). As qubit counts increase, the complexity of mitigating crosstalk, thermal loads, and wiring constraints grows \cite{Preskill2018}. Distributed quantum computing (DQC) has emerged as a scalable paradigm to bypass these monolithic limits. By networking multiple small-scale QPUs via quantum interconnects, a modular, large-scale computational system can be formed, analogous to parallel computing in classical high-performance systems \cite{Cirac1999}.\\
A functional DQC architecture consists of three primary layers: local monolithic QPUs for computation, nonlocal quantum links for inter-processor communication, and a classical control layer for synchronization \cite{Wehner2018}. In this setting, quantum algorithms are decomposed into subcircuits and mapped onto the quantum distributed system. However, the fundamental challenge of DQC lies in the high overhead of nonlocal operations. Executing gates across processor boundaries requires the generation and consumption of shared entanglement, typically Einstein-Podolsky-Rosen (EPR) pairs \cite{Einstein1935}. Whether utilizing quantum state/gate teleportation \cite{Eisert2000, Bennett1993}, entanglement swapping \cite{Huhtanen2026}, and/or direct nonlocal interactions \cite{yuan2026tunablenonlocalzzinteraction}, inter-QPU communication remains the dominant source of noise and decoherence.\\
While research is performed in physical interconnects for DQC, the aforementioned hardware limitations prevent the developing field of distributed quantum algorithms from being experimentally explored \cite{Main2025,baek2025sdqcdistributedquantumcomputing,Awschalom_2021}. This paper addresses the lack of a unified, hardware-aware framework for emulating and benchmarking distributed quantum algorithms on current NISQ devices or classical simulators. Such a framework is critical for industrial applications in fields such as quantum chemistry and financial modeling, where the ability to emulate distributed workloads on NISQ devices or classical systems can significantly accelerate the path to commercial utility \cite{Caleffi2024}.\\
Q-DICE (Quantum Distributed Interconnect Compiler and Emulator), addresses the challenges of emulating a DQC system by introducing a dual-mode framework for distributed emulation using publicly accessible environments, notably classical simulation systems and NISQ-era monolithic quantum hardware. The contributions of this paper are four-fold. We present a novel representation of QPU objects programmatically that supports both classical simulation and hardware-based execution of distributed quantum circuits in Section \ref{sec:topology}. We then derive a methodology for emulating nonlocal quantum link noise on monolithic qubits in Section \ref{sec:noise}. Next, we introduce a constrained mapping algorithm for distributed quantum circuits in Section \ref{sec:transpilation}. Lastly, we validate Q-DICE against both monolithic and real DQC experimental results in Section \ref{sec:results}.\\
This paper is organized as follows. In Section \ref{sec:works}, we analyze the state of the art in distributed quantum compilers and network simulators. Section \ref{sec:method} presents the Q-DICE pipeline and various subroutine methodologies. Finally, Section \ref{sec:results} discusses results of both Q-DICE experimentation and comparative benchmarks.

\section{Related Works}
\label{sec:works}

Research adjacent to Q-DICE spans three broad areas: quantum network simulation, distributed circuit compilation and partitioning, and hardware emulation of distributed quantum systems. 

\subsection{Quantum Network Simulation}

The most mature tools for simulating distributed quantum systems focus on the network and physical layers rather than the algorithmic layer. NetSquid \cite{Coopmans2021}, QuISP \cite{SatohQuISP}, and SimulaQron \cite{Dahlberg2018} are widely adopted platforms for simulating quantum network protocols, entanglement distribution, and repeater architectures at scale. While highly capable in their respective domains, these tools operate at a level of abstraction that is largely disconnected from standard quantum circuit compilation pipelines, making it difficult to evaluate distributed algorithms in a hardware-aware fashion. SimQN \cite{Chen2023} improves on scalability by introducing a modular network-layer simulator, but similarly targets network-level protocol behavior rather than algorithm benchmarking on QPU backends. In general, network-layer simulators tend to focus on probabilistic entanglement generation/distillation rather than higher level network protocols, making them impractical for algorithm evaluation.  A complementary direction is taken by Riera-Sabat et al. \cite{rierasabat2025quantumsimulationnoisyquantum}, who show that NISQ hardware can emulate noisy network channels by shaping native device noise, demonstrating that monolithic hardware is a natural testbed for network processes --- but stops short of addressing distributed circuit mapping under topology constraints.

These tools share two key limitations. First, existing platforms lack support for high-level circuit execution, precluding the assessment of algorithmic performance under realistic nonlocal noise. Second, standard transpilers such as those in Qiskit \cite{JavadiAbhariQiskit}, designed exclusively for monolithic topologies, actively violate distributed boundary constraints when applied to pre-partitioned circuits, as SWAP-based routing does not respect inter-QPU edges. Q-DICE addresses both: a boundary-aware mapping algorithm enforces distributed topology constraints during transpilation, while a noise injection methodology operates at the level of individual nonlocal operations.

\subsection{Distributed Quantum Hardware and Benchmarking}
On the experimental side, Main et al. \cite{Main2025} recently demonstrated distributed quantum computation across two photonically interconnected trapped-ion modules, executing Grover's search algorithm with a 71\% success rate, the first experimental realization of a distributed algorithm comprising multiple non-local two-qubit gates. This landmark result provides both a target benchmark and a noise reference for Q-DICE's validation. The comprehensive DQC survey by Caleffi et al. \cite{Caleffi2024} underscores the absence of unified, hardware-aware benchmarking environments spanning the full DQC stack, directly motivating Q-DICE's cross-layer design. The virtualization of quantum resources explored by Cuomo et al. \cite{Cuomo2021} similarly motivates the need for flexible QPU representations, though that work focuses on multiprogramming rather than teleportation-based distribution.

 Q-DICE is, to our knowledge, the first framework to bridge the gap between abstract distributed quantum algorithms theory and physical implementation. By integrating hardware-faithful topology construction, physically motivated nonlocal noise injection, and distribution-aware circuit mapping, Q-DICE provides a unified system that supports both simulation and physical hardware execution.

\section{System Methodology}
\label{sec:method}
In this section, we discuss the key contribution of this work: an emulation environment for evaluating distributed quantum algorithms. The proposed framework serves as a specialized layer designed to bridge the gap between high-level distributed algorithms and existing monolithic execution environments. It is important to define the scope of this system: it does not function as a standalone simulator or hardware controller. Instead, users provide a predefined distributed circuit, system architecture, and noise information. Then, the Q-DICE system generates two primary elements: (1) a boundary-aware, transpiled circuit mapped to a given connectivity graph, and (2) a corresponding noise-characterized backend model. Users may then leverage these outputs for final execution on physical quantum hardware or classical density matrix, state vector, and tensor network simulations. In this capacity, Q-DICE acts as a ``distribution-aware'' compiler and noise-modeling engine, enforcing the physical limitations of distributed quantum hardware within currently available monolithic or simulated structures.

\subsection{Distributed QPU Topology Engineering}
\label{sec:topology}
\begin{figure}[htbp] 
\centerline{\includegraphics[width=0.8\columnwidth]{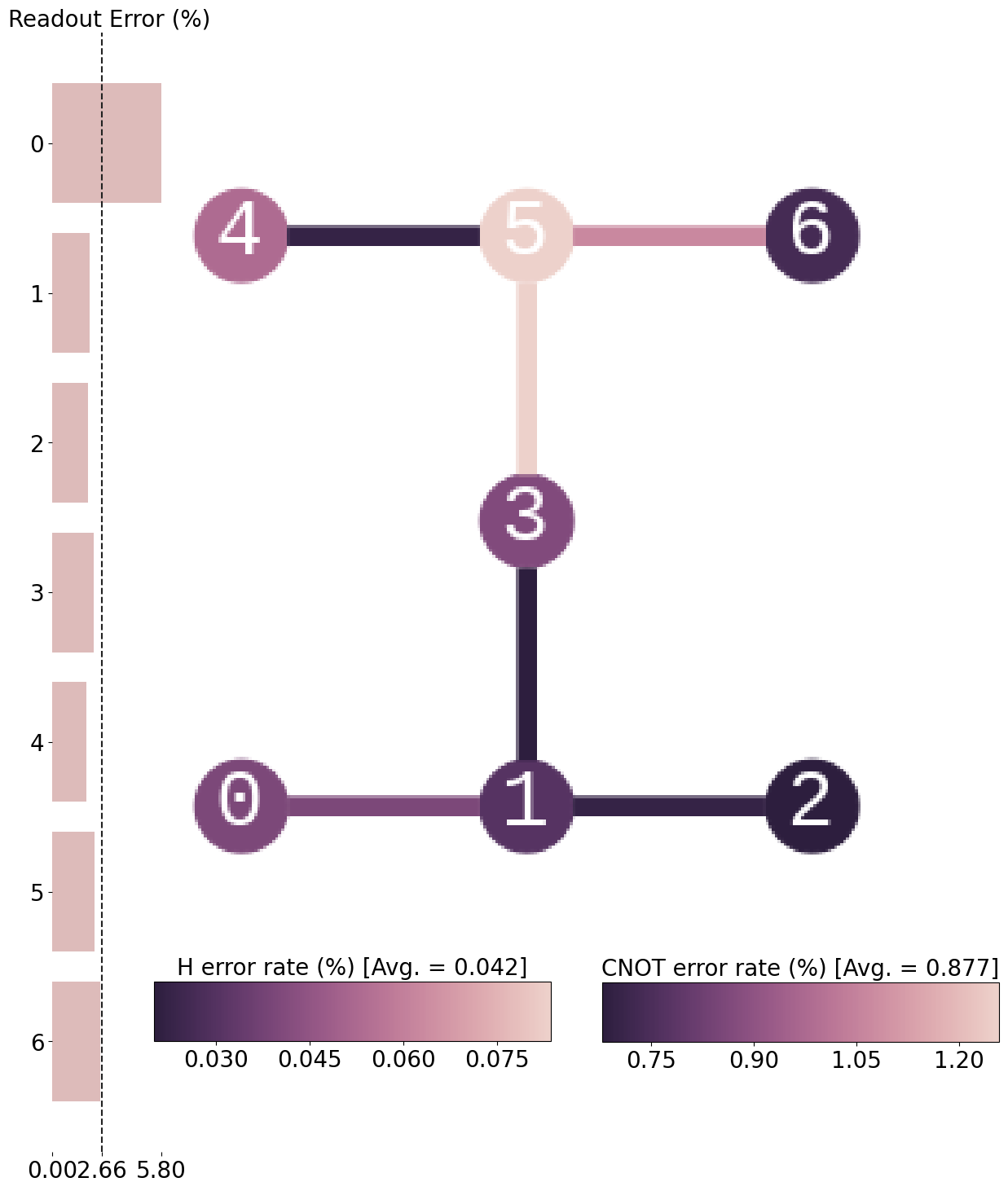}}
\caption{IBM Nairobi monolithic superconducting 27-qubit QPU topology and error map. Graph nodes indicate qubits, with respective colors referring to single gate error rates (in this case for the Hadamard gate). Graph edges represent qubit-qubit connections, with respective colors referring to the error rate of a native two-qubit gate on these qubits. Side bars indicate single-qubit readout error rates.}
\label{fig:errormap}
\end{figure}

In the realm of quantum software engineering, quantum processors are represented as programmatic backend objects, with QPUs modeled as heterogeneous connectivity graphs containing qubit and instruction properties, as shown in Fig.~\ref{fig:errormap}. A backend object encapsulates everything a compiler needs to know about a target device: its qubit count, coupling map (i.e., which pairs of qubits can interact directly), supported native gate set, and per-qubit noise characteristics such as relaxation times $T_1$, dephasing times $T_2$, gate error rates, and readout error rates. These properties are used by the compilation layer to transpile abstract quantum circuits into executable instruction sequences that respect the physical constraints of the target hardware.

Major quantum hardware providers each release backends through their respective software frameworks. IBM Quantum exposes its devices---including the 156-qubit Heron r3 and the 120-qubit Nighthawk processors---through the Qiskit SDK \cite{JavadiAbhariQiskit}, where backends store recent device calibration data. Google Quantum AI exposes its Sycamore and Willow processors through the Cirq framework \cite{Cirq2024}, representing device topology and gate support through \texttt{GridQubit} device objects. Quantinuum exposes its H-series trapped-ion processors through the \texttt{TKET} SDK \cite{Sivarajah2020}, with all-to-all qubit connectivity reflecting the physical properties of ion-trap architectures. In each case, the backend object acts as the hardware abstraction layer between high-level circuit description and low-level physical execution, and it is the primary interface through which noise-aware compilation is performed.\\
These backends are limited to the single-QPU architectures that comprise all current publicly available hardware offerings. As a simulated substitute to running algorithms on physical hardware, backends can also be instantiated within classical simulators---such as Qiskit's \texttt{AerSimulator} \cite{JavadiAbhariQiskit} or Cirq's built-in density matrix simulator \cite{Cirq2024}---that simulate a quantum device's coupling map and noise model using classical computing resources. 
However, both hardware and simulated backends share a fundamental flaw: current implementations are limited to existing monolithic single-QPU systems. This makes them structurally incompatible with distributed quantum architectures, where computation is partitioned across multiple physically separated processors connected by nonlocal quantum links.

Custom backend objects must therefore be constructed to satisfy the unique topologies of a distributed quantum system. Backends have different properties with respect to the mode of execution: hardware implementations are constrained by the monolithic structure and native noise of NISQ QPUs, while simulation-based backends offer the flexibility of controllable custom noise models and larger, user-defined QPU architectures. Q-DICE extends the backend abstraction in both directions. For hardware execution, it partitions an existing monolithic backend into a set of virtual QPUs via QPU \emph{slicing}, cutting edges in the coupling map to enforce inter-QPU boundaries. For simulation, it constructs multi-QPU backends from scratch via QPU \emph{stitching}, connecting user-defined processor topologies through virtual nonlocal links. Both are discussed in detail next.

\subsubsection{QPU Slicing For Hardware Implementation}
\begin{figure}[htbp]
\centerline{\includegraphics[width=0.8\columnwidth]{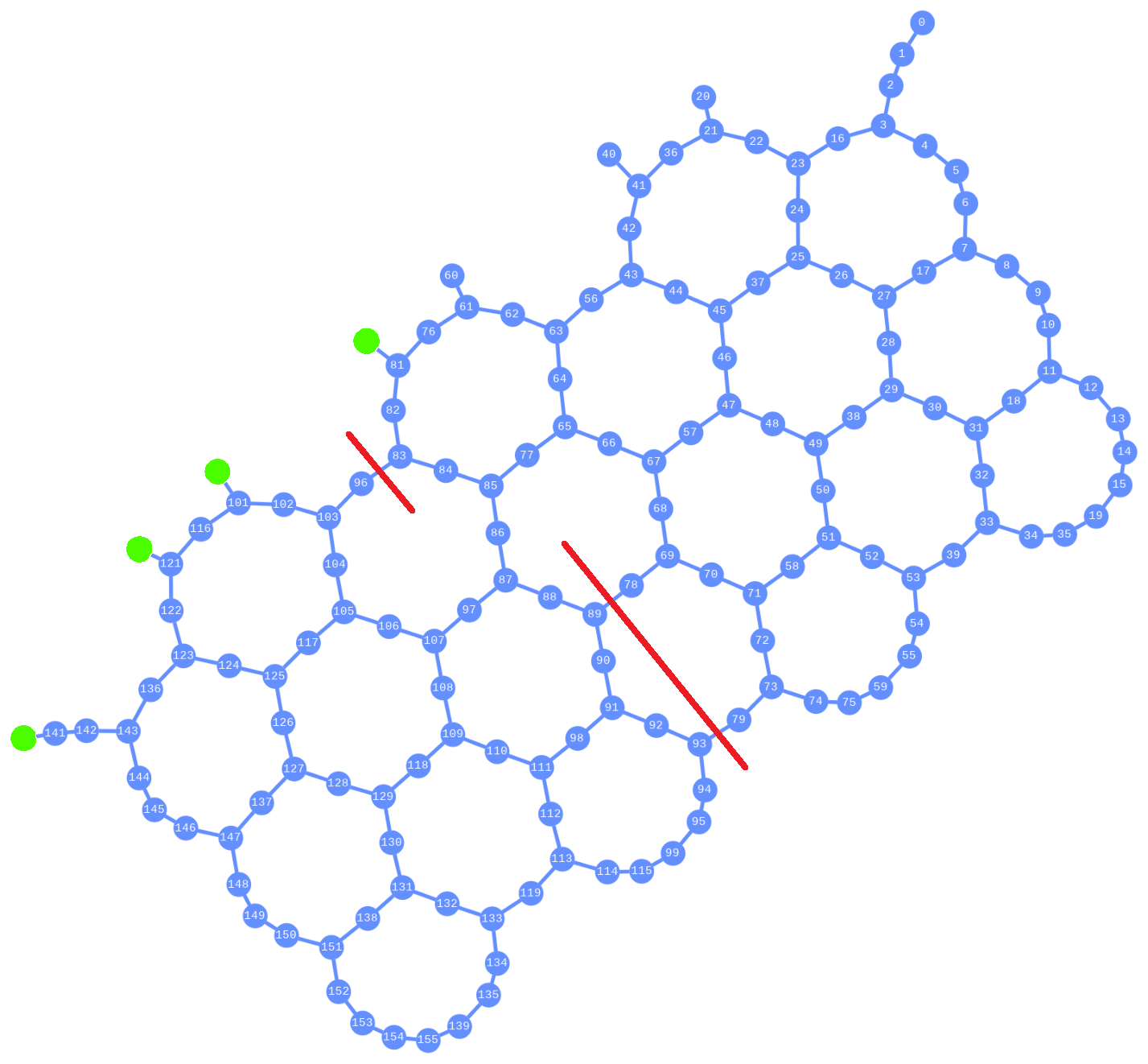}}
\caption{IBM Marrakesh monolithic superconducting Heron 156-qubit QPU coupling map. Red lines indicate user-defined cuts for QPU slicing. Green nodes indicate user-defined communication ancilla qubits for QPU stitching.}
\label{fig:MAP}
\end{figure}

When attempting to implement distributed algorithms on NISQ-era hardware (which is limited to monolithic structures), single QPUs must be partitioned into multiple smaller ``virtual'' QPUs, a process we dub ``QPU slicing''. The user defines which qubit indices are allocated to each virtual QPU, as well as indicating existing qubit connections that should be treated as nonlocal connections. Then, Q-DICE ``slices'' (removes) edges of the monolithic QPU's coupling map (shown in Fig.~\ref{fig:MAP}) so that qubits across virtual processors cannot directly communicate except with the user-specified virtual nonlocal links. 

\begin{figure}[htbp]
\centerline{\includegraphics[width=0.8\columnwidth]{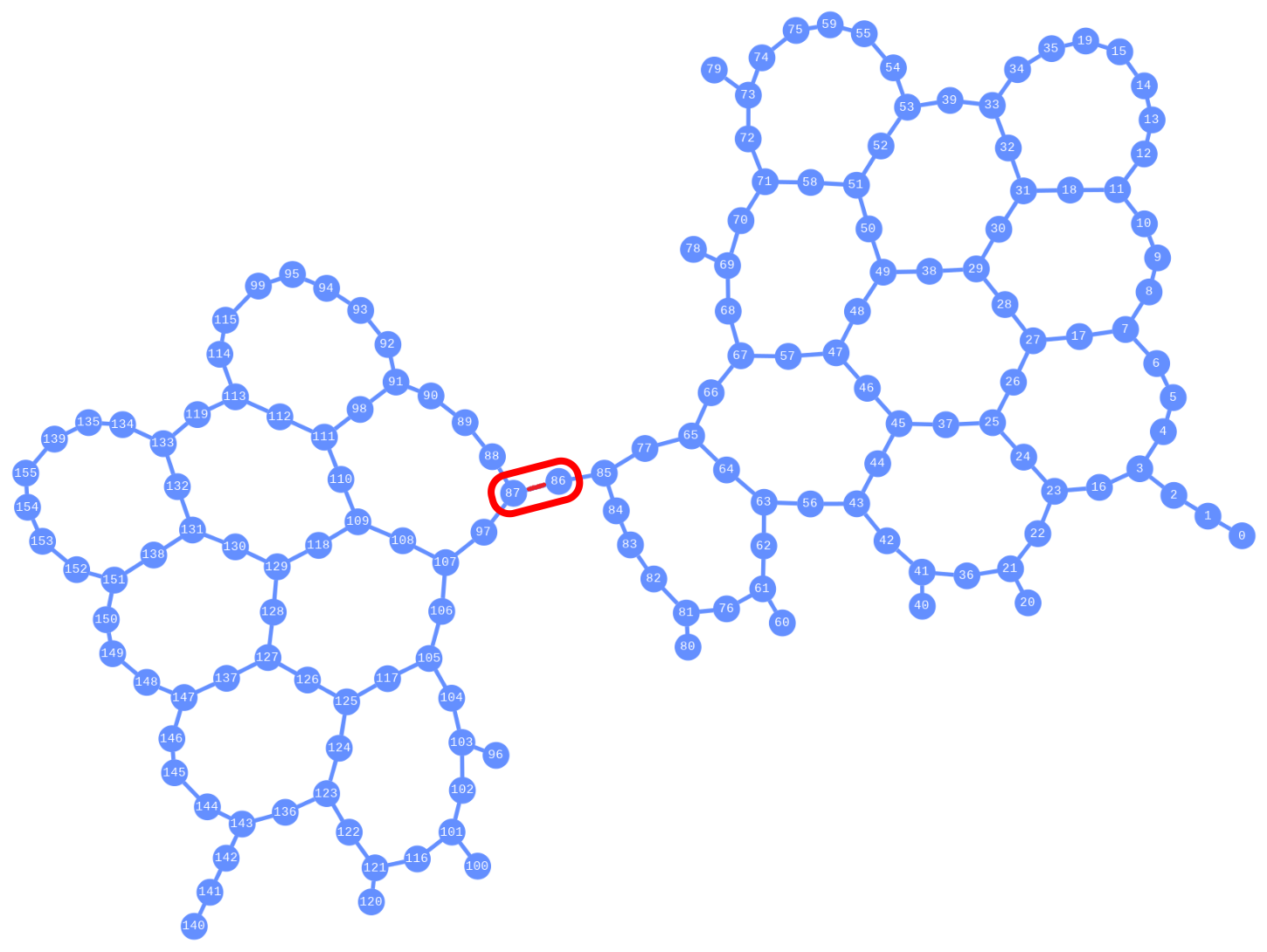}}
\caption{Partitioned IBM Marrakesh processor. Red line indicates virtual nonlocal link, separating two virtual QPUs. We refer to these sparse, modular structures as ``butterfly backends,'' reflecting their uniquely shaped and low-connectivity modular topologies similar to near-term DQC architectures.}
\label{fig:cut_MAP}
\end{figure}

For example, say a user wants to run a circuit distributed across two QPUs and has access to IBM's 156-qubit Heron monolithic QPU. With a reasonably defined system architecture, Q\nobreakdash-DICE will partition the monolithic QPU accordingly into a distributed structure, see Fig.~\ref{fig:cut_MAP} for a visualization of this process. 

It should be noted that qubit nodes connected by virtual nonlocal links function primarily as communication ancilla qubits. Depending on the specific distributed architecture being emulated, these qubits are reserved for inter-processor information transfer and are rarely utilized for local computation.

\subsubsection{QPU Stitching For Classical Simulation}
Simulations gain the benefit of flexibility in QPU architecture, as users are not constrained by fixed monolithic hardware topologies and can dictate their own. 
With respect to Q-DICE, users can define multiple QPU topologies, indicating communication ancillas and QPU-QPU connections. Then, these virtual processors will be ``stitched'' together by adding edges in between communication ancillas, creating a larger distributed system, with the newly created edges acting as nonlocal links. Users also have the option of defining heterogeneous topologies, with processors of multiple different types being stitched together. 

\begin{figure}[htbp]
\centerline{\includegraphics[width=0.8\columnwidth]{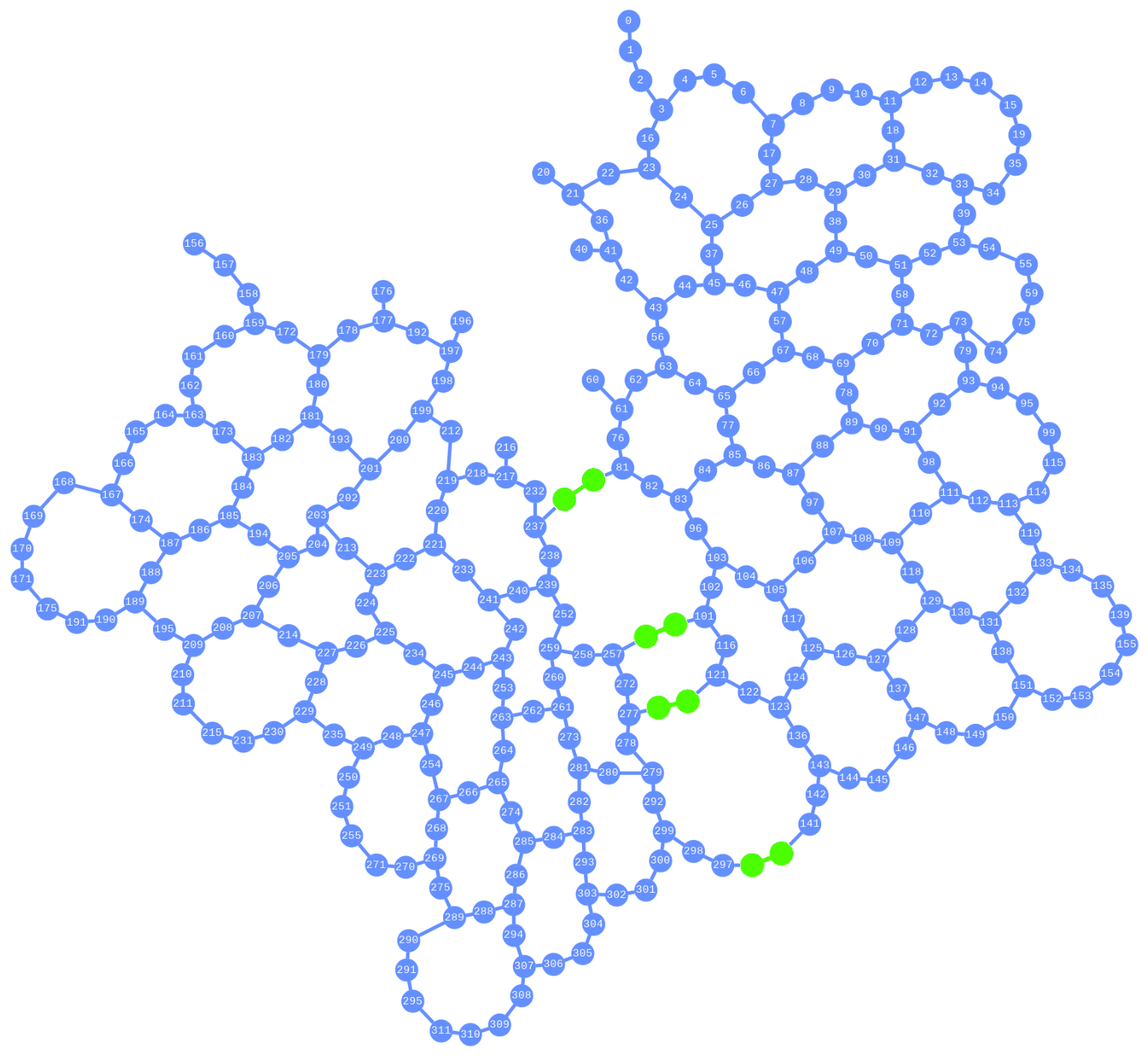}}
\caption{Stitched distributed QPU system. Green nodes indicate communication ancilla qubits, with green edges indicating a virtual nonlocal link between, separating two virtual QPUs.}
\label{fig:stitch_MAP}
\end{figure}

For example, say a user wants to experiment with a large algorithm distributed across two of IBM's 156-qubit Heron monolithic QPUs. By providing the processor's backend information (topology, local noise data, etc.) and defining communication ancillas, Q-DICE will then build a two-QPU system backend that then allows the user to implement and simulate a circuit on. This process is visualized in Fig.~\ref{fig:stitch_MAP}.

\subsection{Nonlocal Noise Characterization}
\label{sec:noise}
To accurately emulate DQC systems, noise must be injected at each nonlocal operation to replicate the error characteristics of physical inter-QPU links. The method of injection differs by execution mode: hardware implementations are limited to probabilistic mixtures of available qubit operations \cite{Niu2024}, while simulation mode applies Kraus operators and error channels directly at the density matrix level for greater accuracy and controllability \cite{Dahlberg2023}. Q-DICE will automatically place the appropriate operators within the compiled circuit and/or backend based on user-defined noise parameters, with local noise drawn from calibration data of the target QPU. Nonlocal noise modeling falls into two categories, corresponding to the two dominant classes of physical distribution schemes.

\subsubsection{Hamiltonian-Mediated Links}
\label{sec:hamil_noise}
Hamiltonian-mediated links are nonlocal channels between QPUs that can implement direct interactions between nonlocal qubits \cite{Cirac1999}. These include schemes such as superconducting microwave resonator buses \cite{yuan2026tunablenonlocalzzinteraction} and phonon-mediated coupling in trapped ion computers \cite{baek2025sdqcdistributedquantumcomputing}. From a quantum information processing perspective, such schemes are dominated by amplitude relaxation and dephasing metrics \cite{Awschalom_2021}. As such, user-defined variables would be $T_1$ (relaxation time, $\mu s- s$), $T_2$ (dephasing time, $\mu s-s$), and $t_{link}$ (remote gate duration, $ns-s$) \cite{Cuomo2021, Caleffi2024}, with time unit magnitudes varying depending on the distribution scheme. \\
In terms of \textit{hardware implementation}, we utilize quantum hardware's natural decoherence by inserting delay instructions, which are generally universal across hardware types. The duration of the delay inserted is 
\[
t_{delay} = \text{max}(0, t_{link} - t_{local gate})
\]
Since the physical QPU must still perform a local gate (acting as a nonlocal one) with time $t_{local gate}$, this consumes a portion of the $t_{link}$ budget. Note that there will be a consistent bottleneck if the user-supplied link time is less than native hardware times, as in that case the hardware operations will always take longer and introduce more noise than needed. This is an unavoidable limitation, however, short- and medium-term DQC link times are still far from local gate times and therefore users will not be affected by this limitation. \\
In terms of \textit{simulation implementation}, the density matrix formalism is used to apply noise operators to the quantum state \cite{Nielsen2010}. Amplitude damping (with time $T_1$) represents the physical process of a qubit losing energy to its environment, with its probability of decay defined as $\gamma = 1- e^{-t_{link}/T_1}$. This is then applied via Kraus operations:
$$E_0 = \begin{pmatrix} 1 & 0 \\ 0 & \sqrt{1-\gamma} \end{pmatrix}, \quad E_1 = \begin{pmatrix} 0 & \sqrt{\gamma} \\ 0 & 0 \end{pmatrix}$$
Phase damping/dephasing (with time $T_2$) represents the loss of quantum phase information without necessarily losing energy. The probability of a phase flip is defined as $\lambda = 1-e^{-t_{link}/T_2}$. The resulting Kraus operators are:
$$E_0 = \begin{pmatrix} 1 & 0 \\ 0 & \sqrt{1-\lambda} \end{pmatrix}, \quad E_1 = \begin{pmatrix} 0 & 0 \\ 0 & \sqrt{\lambda} \end{pmatrix}$$
These operators are then applied every time a virtual nonlocal operation is executed, affecting communication ancillas. This software-driven approach offers high controllability, whereas hardware implementation is constrained by fixed QPU noise profiles (e.g. emulating a superconducting QPU on a photonic QPU may be difficult due to the native noise present in photonic QC).

\subsubsection{Entanglement-Mediated Links}
Entanglement-mediated links are nonlocal channels between QPUs that rely on entanglement generation and subsequent measurements to teleport states and operations \cite{Kimble_2008}. Hardware schemes under this category include silicon T-center interconnects \cite{inc2024distributedquantumcomputingsilicon} and photonically interfaced ion traps \cite{Stephenson_2020}. In both cases, entangled photons are interfered to create a remote link between processors, which can be consumed for nonlocal operations. These approaches are dominated by depolarizing loss in the photonic systems, which from the quantum information processing perspective is essentially unital random loss of information \cite{Wehner2018}. As such, users provide an estimated link error probability $p$ of the emulated link, and $t_{link}$, the nonlocal gate duration. Note that this method of distribution generally requires a sequence of gates for teleportation, as a result $t_{link}$ should be the sum of this entire process.\\
In terms of \textit{hardware implementation}, random loss is implemented using a Monte Carlo approach based on $p$, where, if it's determined an error occurred, a random Pauli gate is applied, each of which ($\{X,Y,Z\}$) then has probability $\frac{1}{3}$ of being applied. This represents a stochastic randomization of the transmitted state, forcing the system towards a maximally mixed state as nonlocal link fidelity decreases. Such an approach is similar to Pauli twirling in building error models \cite{Geller_2013}. \\ 
In terms of \textit{simulation implementation}, a two-qubit depolarizing channel is used as the standard model for these links \cite{Nielsen2010}. If a remote gate is performed with an error probability $p$, the state of the two qubits $\rho$ evolves as:
$$\mathcal{E}(\rho) = (1-p)\rho + p\frac{I}{4}$$
This can also be written in the Kraus representation, using the Pauli basis:
$$\mathcal{E}(\rho) = (1 - p)\rho + \frac{p}{15} \sum_{i\neq I\otimes I}^{15} P_i \rho P_i$$
Similar to Section \ref{sec:hamil_noise}, these are applied every time a nonlocal operation takes place.

\subsection{Boundary-Aware Circuit Mapping}
\label{sec:transpilation}
Modern circuit compilation/transpilation algorithms are built exclusively for optimization on monolithic hardware. Standard qubit routing algorithms use SWAP mapping to overcome challenges in limited QPU topologies, however, this automatic gate placement tends to break the virtual boundaries between QPUs introduced in QPU stitching/slicing. Additionally, qubit layout algorithms generally do not support preplaced qubits and will change layouts for circuit optimization, which generally breaks communication ancilla constraints. This then leads to nonlocal operations happening on local links, and vice versa, essentially ruining the distributed architecture.
Lastly, gate optimizations and insertions usually result in local operations happening illegally across virtual nonlocal links. In summary, a regular transpiler tends to turn distributed circuits into optimized local ones. \\
Q-DICE addresses these shortcomings through a constrained layout selection algorithm that enforces distributed boundary conditions when mapping circuits to backends, as outlined in Alg.\ref{alg:mapping}. The algorithm operates in three phases per iteration, repeating a user-supplied number of iterations $N$, returning the minimum-depth circuit found in any iteration as the final output.

\begin{algorithm}
\label{alg:mapping}
\DontPrintSemicolon
\SetKwInOut{Input}{Input}\SetKwInOut{Output}{Output}
\Input{Virtual circuit $C$, set of virtual partitions $\{vQPU_k\}$, set of communication ancillas $A$, port mapping $A \to P$, restricted coupling map $G_{restricted}$, number of iterations $N$}
\Output{Optimized transpiled circuit $C_{best}$ and layout $\pi_{best}$}

$depth_{best} \leftarrow \infty$\;
$\pi_{best} \leftarrow \text{None}$\;

\For{$i \leftarrow 1$ \KwTo $N$}{
    \ForEach{$a \in A$}{
        Fix $a$ to physical port $p \in P$ at boundary of $vQPU_{home(a)}$\;
    }
    
    \ForEach{partition $vQPU_k$}{
        Identify unassigned circuit qubits $V_{rem} \subset vQPU_k$\;
        Identify available qubits $Q_{avail}$ in physical subgraph of $vQPU_k$\;
        Randomly map $V_{rem} \to Q_{avail} \setminus P$ to form partial layout $\pi_i$\;
    }
    
    $C_{temp} \leftarrow \text{transpile}(C, \text{initial\_layout}=\pi_i, \text{coupling\_map}=G_{restricted})$\;
    
    \If{$\text{depth}(C_{temp}) < depth_{best}$}{
        $depth_{best} \leftarrow \text{depth}(C_{temp})$\;
        $C_{best} \leftarrow C_{temp}$\;
        $\pi_{best} \leftarrow \pi_i$\;
    }
}
\Return $C_{best}, \pi_{best}$\;
\caption{Distributed-Aware Circuit Mapping}
\end{algorithm}

We define a port qubit $p \in P$ as a physical qubit located on the boundary of a QPU chip capable of inter-chip communication, and the virtual QPU ($vQPU_k$) as the range of qubits constrained in a partition assigned by the user (see Section \ref{sec:topology}).\\
Prior to qubit assignments, each communication ancilla qubit $a \in A$ is deterministically fixed to a distinct physical port qubit, at the boundaries of their assigned virtual QPU, $\mathrm{vQPU}_{home(a)}$ as dictated by the user. 
This ensures every virtual nonlocal link corresponds to a valid edge in the restricted backend coupling map $G_{restricted}$. Remaining computational qubits are randomly assigned to free physical qubits, subject to the hard constraint that each qubit must be placed within the physical subgraph of its home virtual QPU, generating a layout $\pi$. This ensures that all local gates within a partition operate on physically co-located qubits, respecting the local connectivity of the backend's partitions. The random component introduces layout diversity across iterations, allowing the search to explore placements that may yield lower circuit depth. Finally, any standard monolithic transpiler, such as those found in existing quantum software stacks, may be invoked with the fixed layout $\pi$ and a restricted coupling map $G_\text{restricted}$ from which inter-QPU edges have been removed except for the user-designated virtual nonlocal links. By restricting the coupling map in this way, the transpiler is prevented from inserting operations across partition boundaries, keeping the distributed architecture intact. The transpiled circuit, $C_{temp}$, is then evaluated by circuit depth, and retained as the current best solution if it improves on prior iterations. 

The algorithm is inherently a constrained random search and does not guarantee globally optimal layout selection. However, for the class of distributed circuits considered in this work---those designed for near-term DQC architectures---the search space of valid layouts is significantly reduced relative to the unconstrained monolithic case, and near-optimal solutions are found reliably within a modest number of iterations $N$, with time complexity $\mathcal{O}(N \cdot T_{\text{transpile}})$ and space complexity $\mathcal{O}(|Q| + |E|)$, where $|E|$ is the coupling map edge count.  Future work will explore advanced mapping algorithms, mainly graph-theoretic layout methods, such as subgraph isomorphism or hypergraph-based placement, to improve scalability to larger distributed systems.

\section{Results}
\label{sec:results}

This section presents experimental and simulation results validating Q-DICE across four classes of distributed quantum circuits. Experiments were selected to span foundational DQC primitives, the primary available experimental benchmark, and novel algorithm/architecture scenarios beyond existing validated results. All hardware experiments were conducted on IBM's 156-qubit Marrakesh Heron r2 superconducting processor accessed via the IBM Quantum cloud platform \cite{JavadiAbhariQiskit}. Hardware-mode distributed experiments used QPU slicing to partition the monolithic Marrakesh backend into virtual QPUs, with user-specified cuts defining inter-QPU boundaries and virtual nonlocal links. Simulation-mode experiments used QPU stitching to construct custom multi-QPU backends with controllable noise models, as described in Section \ref{sec:method}. For each experiment, Q-DICE noise parameters were set in accordance with values reported in the corresponding experimental reference, as detailed in each subsection below.

A general observation across all experiments is that superconducting hardware implementations of Q-DICE emulate Hamiltonian-mediated link noise more faithfully than entanglement-mediated link noise. This is a direct consequence of the physical correspondence between the two: amplitude damping and dephasing channels, which dominate Hamiltonian-mediated links, are natively present in superconducting NISQ hardware and can be induced controllably via delay instructions. Depolarizing loss, which dominates entanglement-mediated (photonic) links, is less naturally replicated on superconducting hardware, requiring stochastic Pauli injection whose fidelity is bounded by the native gate error floor of the device. Simulation mode suffers from neither limitation, as noise is applied analytically at the density matrix level.

\subsection{Gate Teleportation}
\label{sec:gate_tele}

\begin{figure} 
\begin{center}
\begin{adjustbox}{width=\columnwidth}
\begin{quantikz}
\lstick{Node 1: $q_{data}$} && & \ctrl{1}&&&& \gate{Z} & \\
\lstick{Node 1: $q_{comm}$}  & \gate{H}\gategroup[wires=3,steps=2,style={dashed}]{Nonlocal Link} & \ctrl{2} & \targ{} & \meter{}  \wire[d][2]{c} \\
\wave{}&&&&&&&&&& \\
\setwiretype{q}
\lstick{Node 2: $q_{comm}$}  & & \targ{} & & \gate{X} & \ctrl{1} \gategroup[wires=2,steps=1,style={dashed},label style ={label position = below, anchor=north, yshift=-0.2cm}]{Teleported Gate} & \gate{H} & \meter{} \wire[u][3]{c}\\
\lstick{Node 2: $q_{data}$} &&&&& \targ{} &&&
\end{quantikz}
\end{adjustbox}
\end{center}
\caption{Gate teleportation circuit. The dashed box labeled ``Nonlocal Link''
encloses the entanglement generation and distribution subcircuit consumed by
the teleportation. The dashed box labeled ``Teleported Gate'' encloses the
operation applied at Node 2 conditioned on classical measurement outcomes
from Node 1.}
\label{fig:gate_tele_circuit}
\end{figure}
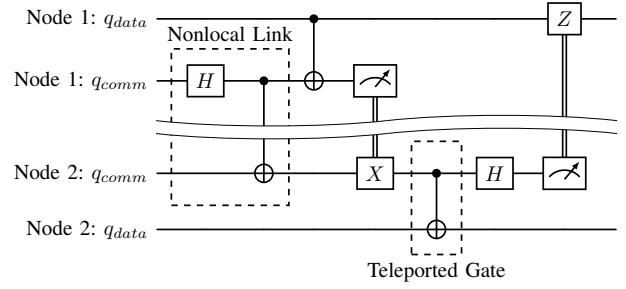

\begin{figure}
    \centering
    \includegraphics[width=0.4\columnwidth]{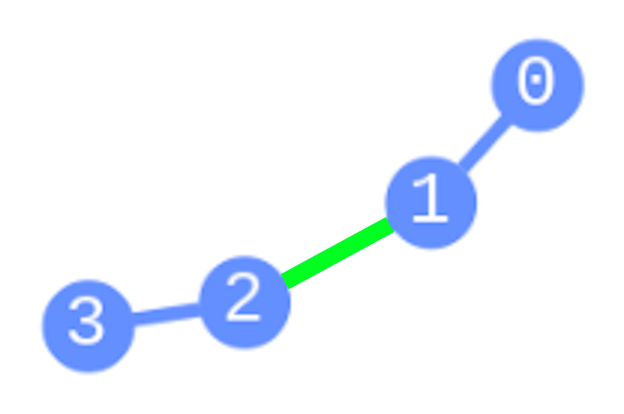}
    \caption{Two-node simulated backend for gate teleportation experiment.
    The green edge indicates a virtual nonlocal link between communication ancillas.}
    \label{fig:gate_tele_backend}
\end{figure}

\begin{figure} 
\centerline{\includegraphics[width=0.5\textwidth]{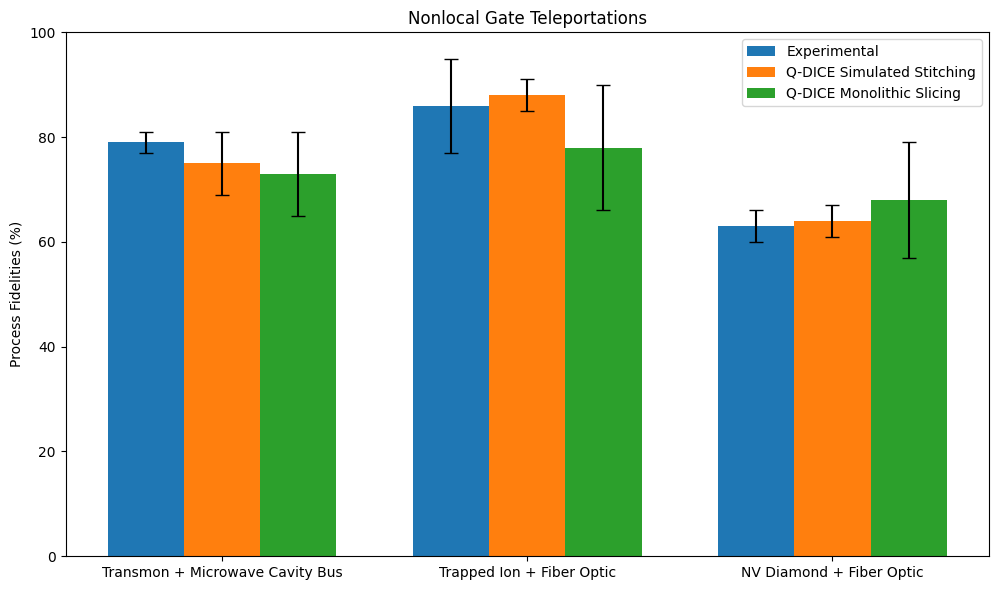}}
\caption{Teleported gate fidelity across Q-DICE emulation modes and
experimental references \cite{Chou_2018}, \cite{Main2025}, \cite{iuliano2026unconditionallyteleportedquantumgates}. Error bars on experimental values reflect
run-to-run variation averaged across repeated experiments.} 
\label{fig:gate_tele_plot} 
\end{figure}

Gate teleportation is the fundamental primitive enabling non-local two-qubit
operations in entanglement-mediated distributed quantum systems. In this
experiment, a controlled-Z (CZ) gate is teleported between two nodes using
a shared entangled pair, as illustrated in Fig.~\ref{fig:gate_tele_circuit}.
The two-node backend used for simulation is shown in Fig.~\ref{fig:gate_tele_backend}.
The teleportation fidelity is measured via quantum process tomography and
compared against three experimental references spanning both link paradigms (seen in Fig.~\ref{fig:gate_tele_plot}).

\subsubsection{Hamiltonian-Mediated Reference}

Chou et al.\ demonstrated deterministic teleportation of a quantum gate
between two logical qubits in a superconducting circuit system connected
via a microwave cavity bus, reporting a teleported gate process fidelity of
approximately 79\%. This system is Hamiltonian-mediated: the microwave
resonator bus mediates a direct qubit-qubit interaction between the two
modules, making its noise profile dominated by amplitude relaxation and
dephasing. For the Q-DICE Hamiltonian-mediated noise model, we set
$T_1 = 50\,\mu\text{s}$ and $T_2 = 30\,\mu\text{s}$ consistent with
the reported transmon coherence regime of the experiment, and
$t_{\text{link}} = 2\,\mu\text{s}$, consistent with the microwave-mediated
gate duration reported in that work. These parameters yield amplitude damping
probability $\gamma = 1 - e^{-t_{\text{link}}/T_1} \approx 0.039$ and
dephasing probability $\lambda = 1 - e^{-t_{\text{link}}/T_2} \approx 0.065$.
The superconducting Marrakesh hardware is well-suited to emulating this
noise regime, as native decoherence in the device is physically identical
in character to the target noise, and delay instructions accurately reproduce
the $t_{\text{link}}$ budget within the hardware mode.

\subsubsection{Entanglement-Mediated Reference}
Main et al.\ demonstrated deterministic gate teleportation between two
photonically interconnected trapped-ion modules, achieving a CZ gate process
fidelity of 86\%. This system is entanglement-mediated: remote entanglement
is generated via photon interference across an optical fiber link and consumed
to execute the teleportation. Its noise profile is dominated by depolarizing
loss in the photonic channel. For the Q-DICE entanglement-mediated noise model,
we set the link error probability $p = 0.14$, derived directly from the
reported 86\% gate fidelity, and $t_{\text{link}}$ equal to the full
teleportation protocol duration including entanglement generation and
feed-forward classical communication. The two-qubit depolarizing channel
with $p = 0.14$ is applied at each instance of the nonlocal CZ operation
in both simulation and hardware modes. As expected, the simulation mode
achieves closer agreement with the experimental result than the hardware
mode for this reference, since depolarizing noise is less faithfully
reproduced by native superconducting decoherence than by the analytical
density matrix channel.

\subsubsection{Entanglement-Mediated Reference}

Iuliano et al.\ recently demonstrated unconditional gate teleportation between nitrogen-vacancy diamond qubit registers connected via optical fiber. This experiment represents a distinct physical implementation of the same entanglement-mediated paradigm as Main et al. \cite{Main2025}, with different error characteristics reflecting the NV-center photonic interface. Q-DICE noise parameters for this reference were set to match the reported link fidelity (76.5\%) of the NV-photon entanglement generation step, with the corresponding depolarizing error rate ($p=23.5\%$) applied per teleportation instance.

\subsection{GHZ State Generation} 
\label{sec:ghz}

\begin{figure}
    \centering
    \includegraphics[width=0.4\columnwidth]{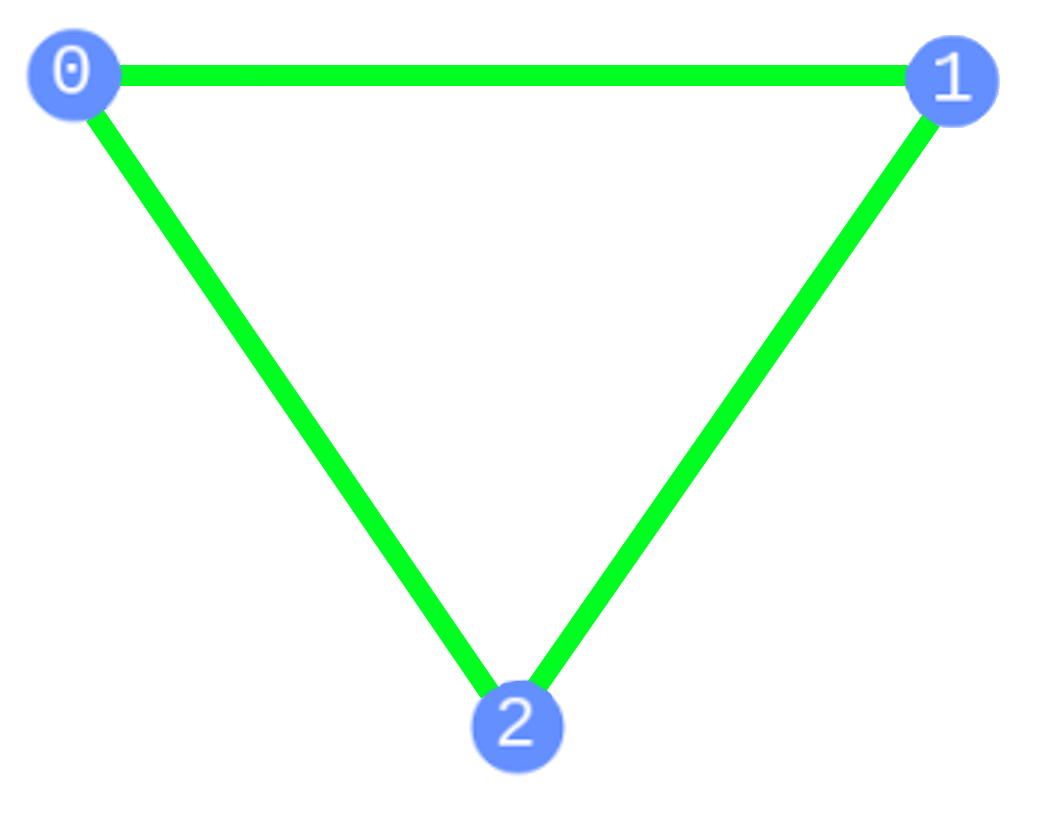}
    \caption{Three-node simulated backend for GHZ state generation.
    Green edges indicate virtual nonlocal links between communication ancillas.
    Each node contains one computational qubit and one communication ancilla.}
    \label{fig:GHZ_backend}
\end{figure}

The generation of Greenberger-Horne-Zeilinger (GHZ) states across multiple
distributed nodes is a canonical benchmark for multipartite entanglement in
quantum networks, and requires sequential nonlocal operations across more
than one inter-QPU link. This experiment was selected because three-node
distributed GHZ state generation has been experimentally realized across
multiple hardware platforms and link paradigms \cite{Pompili_2021,
yan2025quantumsecretsharingtriangular, Tsujimoto_2018}, providing a
diverse set of noise references against which Q-DICE can be validated.
The three-node simulated backend is shown in Fig.~\ref{fig:GHZ_backend}.

\begin{figure}
\centerline{\includegraphics[width=0.5\textwidth]{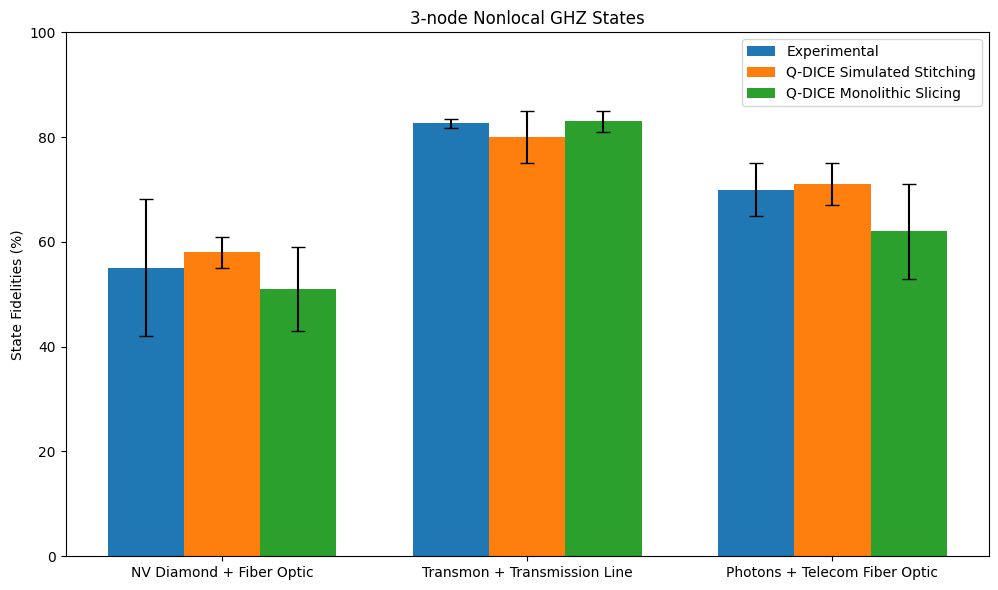}}
\caption{GHZ state fidelity across Q-DICE emulation modes and experimental
references spanning entanglement-mediated \cite{Pompili_2021}, \cite{Tsujimoto_2018} and Hamiltonian-mediated link \cite{yan2025quantumsecretsharingtriangular}
paradigms.}
\label{fig:ghz_plot}
\end{figure}

\subsubsection{Entanglement-Mediated Reference}

Pompili et al.\ demonstrated the realization of a three-node quantum network of remote NV-center qubits connected via optical fiber, generating GHZ states across all three nodes. The experiment is entanglement-mediated, with the photonic link between each pair of nodes subject to depolarizing loss. We set the Q-DICE inter-node depolarizing error probability to $p = 0.20$ per link, consistent with the reported per-link entanglement fidelity in that work, applied at each nonlocal entangling operation in the GHZ generation circuit. The three-node backend is constructed via QPU stitching in simulation mode, with each node instantiated as a homogeneous two-qubit virtual QPU (one computational qubit, one communication ancilla) and connected by two virtual nonlocal links arranged in a linear chain topology.

\subsubsection{Hamiltonian-Mediated Reference}

Yan et al.\ demonstrated a three-node GHZ state taking place in a triangular network of three superconducting transmon qubits connected by microwave transmission lines, requiring multipartite entanglement generation across all three nodes. This is a Hamiltonian-mediated system, with amplitude relaxation and dephasing as the dominant noise sources. Q-DICE noise parameters were set to
$T_1 = 80\,\mu\text{s}$, $T_2 = 50\,\mu\text{s}$, and
$t_{\text{link}} = 1.5\,\mu\text{s}$, consistent with typical superconducting
transmon coherence and microwave interaction timescales. This reference is
the most physically similar to the IBM Marrakesh hardware used in the
hardware execution mode, and consequently the hardware-mode Q-DICE results
show the closest agreement with this experimental reference of all experiments
in this section.

\subsubsection{Entanglement-Mediated Reference}

Tsujimoto et al.\ demonstrated three-qubit GHZ state generation using asynchronous telecom photon pair sources over optical fiber. Q-DICE models this system as an entanglement-mediated link with a depolarizing error probability ($p=16\%$) derived from the reported entanglement swapping fidelity (84\%), applied at the swapping operation that effectively creates the third node's entanglement with the rest of the network.

\subsection{Distributed Grover's Search}
\label{sec:grovers}

\begin{figure} 
\centerline{\includegraphics[width=0.8\columnwidth]{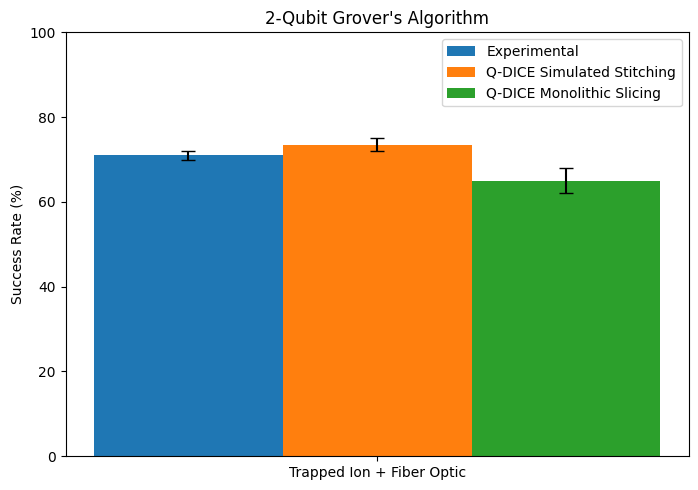}}
\caption{Success rate of distributed 2-qubit Grover's search algorithm
across Q-DICE emulation modes, compared against the experimental result
of Main et al.\ \cite{Main2025}. The worst-case deviation between Q-DICE
simulation and the experimental result is 4\%.}
\label{fig:grovers}
\end{figure}

Distributed Grover's search constitutes the primary quantitative validation
benchmark for Q-DICE. Main et al.\ \cite{Main2025} demonstrated the first
implementation of a distributed quantum algorithm comprising multiple
non-local two-qubit gates, executing a 2-qubit Grover's search algorithm
across two photonically interconnected trapped-ion modules and reporting a
71\% success rate. The experimental system is entanglement-mediated, with
each non-local CZ gate consuming one instance of heralded remote entanglement
across the optical fiber link. In this experiment, Q-DICE is tasked with
reproducing this result as closely as possible using both simulation and
hardware execution modes, with no parameter fitting performed after the
initial noise characterization.

The Q-DICE noise model for this experiment is identical to that used in
the gate teleportation experiment against the same reference
(Section \ref{sec:gate_tele}): a two-qubit depolarizing channel with
$p = 0.14$ applied at each instance of a nonlocal CZ gate, and a link
duration $t_{\text{link}}$ matching the full entanglement generation and
feed-forward protocol of that system. The distributed Grover's circuit
comprises two nonlocal CZ gates, so the depolarizing channel is applied
twice in sequence over the course of circuit execution. The boundary-aware
mapping algorithm described in Section \ref{sec:transpilation} is used to
ensure that the two nonlocal gates are routed correctly across the virtual
inter-QPU link without SWAP insertion.

The Q-DICE simulation mode achieves a worst-case fidelity deviation of 4\%
from the experimental result of Main et al., demonstrating that the
density-matrix-level noise model accurately captures the dominant error
mechanisms of the photonic link system. The hardware execution mode shows
a larger deviation, consistent with the general observation that
superconducting native decoherence does not faithfully replicate photonic
depolarizing loss for entanglement-mediated links. Nevertheless, the
hardware result remains within a physically interpretable margin, confirming
that Q-DICE hardware mode provides a conservative lower bound on expected
distributed circuit fidelity for superconducting platforms.

\subsection{Distributed Deutsch-Jozsa Algorithm}
\label{sec:dj}

\begin{figure}
\begin{center}
\begin{adjustbox}{width=\columnwidth}
\begin{quantikz}[row sep=0.1cm]
\lstick{\ket{0}} & \gate{H} & \ctrl{4} & \qw      & \qw      & \qw      & \gate{H} & \meter{} \\
\lstick{\ket{0}} & \gate{H} & \qw      & \ctrl{3} & \qw      & \qw      & \gate{H} & \meter{} \\
\lstick{\ket{0}} & \gate{H} & \qw      & \qw      & \ctrl{2} & \qw      & \gate{H} & \meter{} \\
\lstick{\ket{0}} & \gate{H} & \qw      & \qw      & \qw      & \ctrl{1} & \gate{H} & \meter{} \\
\lstick{\ket{1}} & \gate{H} & \targ{}  & \targ{}  & \targ{}  & \targ{}  & \qw      & \qw      
\end{quantikz}
\end{adjustbox}
\end{center}
\caption{Balanced 4-qubit Deutsch-Jozsa circuit. In the distributed
implementation, the ancilla qubit (bottom) is placed on Node 2, while
the query qubits are placed on Node 1, requiring one nonlocal CNOT per
query qubit via the inter-QPU link.}
\label{fig:DJ_circuit}
\end{figure}
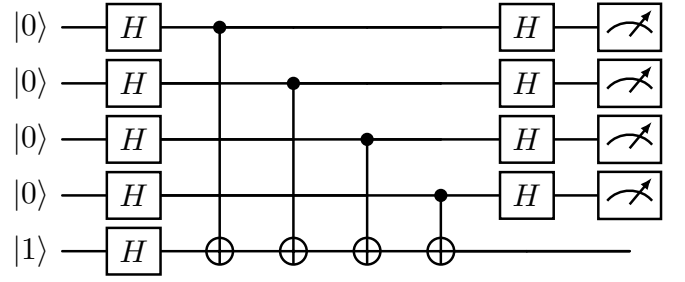

\begin{figure}
    \centering
    \includegraphics[width=0.8\columnwidth]{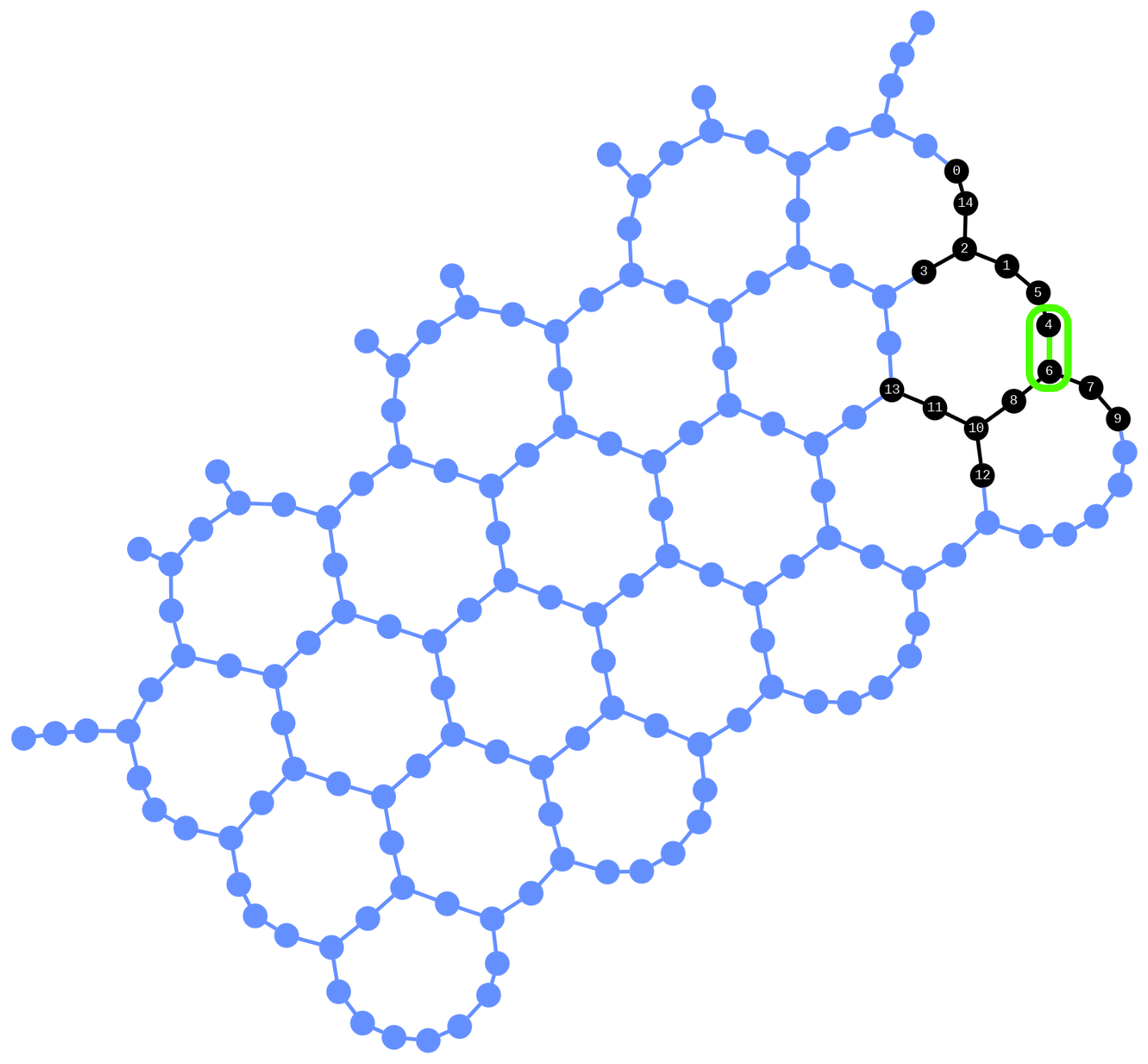}
    \caption{Circuit map of the distributed 15-qubit balanced
    Deutsch-Jozsa algorithm on the IBM Marrakesh Heron r2 QPU. The
    highlighted green edge indicates the virtual nonlocal link, with
    the QPU sliced into two virtual QPUs of approximately equal qubit
    count, each holding half the total circuit qubits.}
    \label{fig:DJ_circuit_map}
\end{figure}

\begin{figure}
    \centering
    \includegraphics[width=\linewidth]{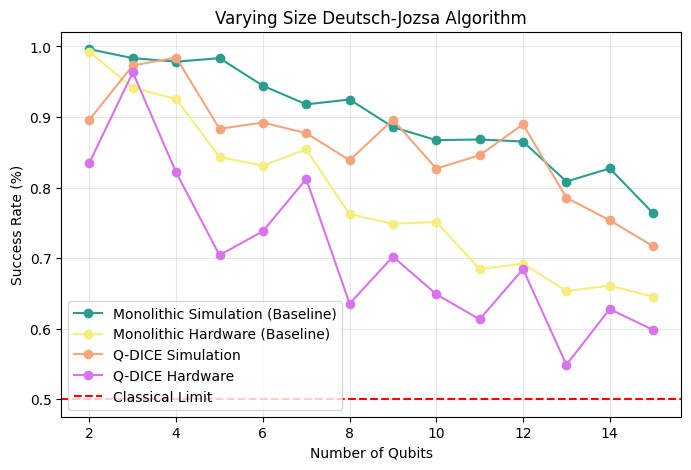}
    \caption{Success rate of the balanced Deutsch-Jozsa algorithm across
    four execution modes on the IBM Marrakesh Heron r2 QPU: the two non-distributed baselines of monolithic
    hardware and monolithic simulated, as well as the two distributed implementations Q-DICE simulation and Q-DICE hardware. The dashed
    horizontal line at fidelity $= 0.5$ indicates the success rate for random guessing; since this is the optimal strategy for a classical algorithm with single oracle call, any quantum
    implementation with success rate above this line shows a quantum
    advantage.}
    \label{fig:DJ_plot}
\end{figure}

The Deutsch-Jozsa algorithm was selected as the final hardware-mode benchmark for two reasons: it enables a direct comparison between distributed and non-distributed execution on the same physical hardware, isolating the overhead of QPU slicing and virtual nonlocal link noise from the underlying hardware noise floor; and its star-topology oracle, where all CNOT gates target a single ancilla, means nonlocal operations scale linearly and independently, making it more robust to distribution than circuits requiring all-to-all connectivity such as the quantum Fourier transform. We restrict experimentation to the balanced oracle exclusively, as the constant oracle introduces no inter-QPU operations and is irrelevant for distributed benchmarking.

The experiment was conducted on the IBM Marrakesh Heron r2 processor using Q-DICE's QPU slicing mode. The 156-qubit monolithic backend was sliced into two virtual QPUs of approximately equal qubit count, with a single virtual nonlocal link designated between two adjacent physical qubits at the cut boundary, as shown in Fig.~\ref{fig:DJ_circuit_map}. A 15-qubit balanced Deutsch-Jozsa instance was compiled using our boundary-aware mapping algorithm, with the circuit partitioned evenly in half: approximately half the query qubits are placed on Node 1, the other half on Node 2, and the ancilla qubit is similarly assigned to one of the two nodes. This halving partition ensures that oracle CNOT gates between query qubits on Node 1 and the ancilla on Node 2---and vice versa---each constitute a nonlocal operation across the virtual link, exercising the full Q-DICE nonlocal compilation and noise injection pipeline. The Q-DICE Hamiltonian-mediated noise model was applied at each nonlocal CNOT operation, with $T_1=179.71 \mu s$, $T_2=88.38\mu s$, and $t_{\text{link}} = 70ns$ values taken from the IBM Marrakesh device calibration data at the time of execution.

Four execution conditions were measured and compared: (1) monolithic hardware baseline, where the circuit is transpiled without any distribution constraints and executed on the full Marrakesh backend; (2) monolithic simulation baseline, using Qiskit's \texttt{AerSimulator} with the Marrakesh noise model but again no distribution; (3) Q-DICE distributed simulation, using the sliced backend with the Hamiltonian-mediated noise channel applied at nonlocal operations in density matrix simulation; and (4) Q-DICE distributed hardware, using the sliced Marrakesh backend with physical execution. The results are shown in Fig.~\ref{fig:DJ_plot}. The dashed baseline at fidelity $= 0.5$ marks the classical random guessing bound: any execution mode reaching this threshold provides no computational advantage over the optimal single-oracle-call classical strategy (i.e., random guessing), so maintaining success rates meaningfully above this line is the minimum criterion for demonstrating quantum utility.

The monolithic simulation baseline achieves the highest success rate, incurring no nonlocal overhead. Q-DICE distributed simulation closely tracks the distributed hardware result, confirming the noise model captures the dominant fidelity loss sources. The consistent degradation observed between distributed hardware and simulation as circuit size grows reflects accumulation of local NISQ decoherence across the larger circuit depth imposed by distributed mapping, a fundamental hardware-mode limitation that simulation avoids by applying noise analytically.

\subsection{Implementing Novel System Architectures}
\label{sec:ror}

\begin{figure}
    \centering
    \includegraphics[width=0.6\columnwidth]{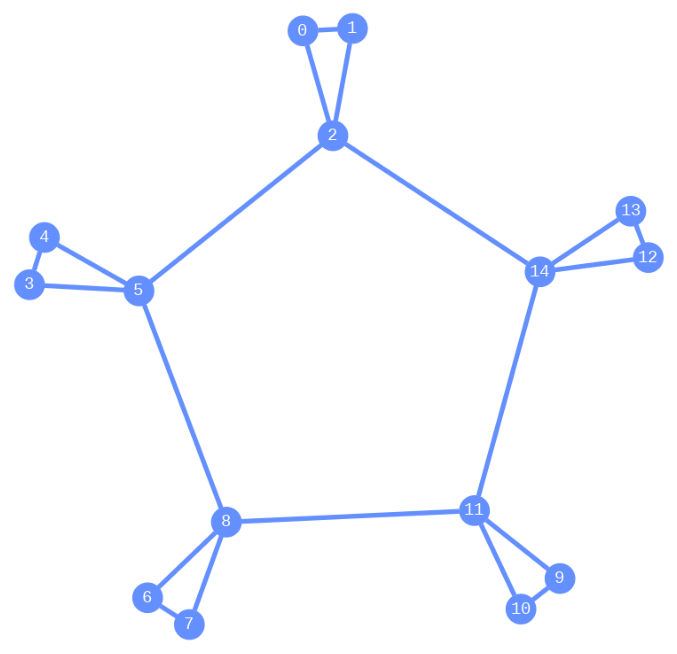} 
    \caption{Ring-of-rings simulated backend. Five triangular QPU modules,
    each comprising two computational qubits and one communication ancilla
    (highlighted), are connected in a ring topology via inter-module nonlocal
    links between adjacent communication ancillas.}
    \label{fig:ror_backend}
\end{figure}

\begin{figure}
    \centering
    \includegraphics[width=\linewidth]{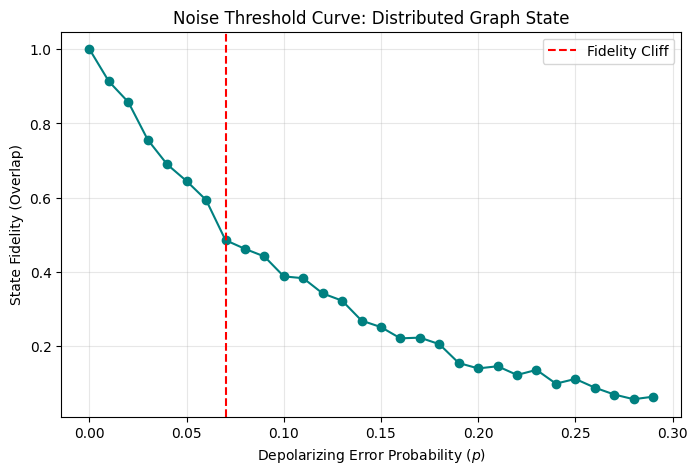}
    \caption{Graph state fidelity on the ring-of-rings architecture across
    multiple entanglement-mediated link error rates $p$, demonstrating
    Q-DICE's utility for noise-threshold analysis in novel distributed
    architectures. The red line indicates the depolarizing error probability at which noise overwhelms the graph state's entanglement.}
    \label{fig:ror_plot}
\end{figure}

A key motivation for Q-DICE is the ability to evaluate distributed quantum algorithms on novel architectural topologies. To demonstrate this capability, we construct a ring-of-rings backend in simulation mode using QPU stitching, and use it to benchmark graph state generation---a resource state foundational to measurement-based quantum computing and distributed quantum protocols \cite{Nielsen2010}.

The ring-of-rings architecture consists of five triangular QPU modules, each comprising three qubits: two computational qubits and one communication ancilla located at the apex of the triangle. The five modules are connected in a ring by virtual nonlocal links between adjacent modules' communication ancillas, forming a 15-qubit distributed system with a ring-of-rings connectivity structure. This topology is of practical interest as a candidate modular architecture for several reasons. The intra-module triangular connectivity provides richer local entanglement than a linear chain, directly analogous to the multi-qubit register structure proposed in the foundational MUSIQC modular architecture \cite{Monroe2014}, where small trapped-ion registers form the elementary computational units of a larger photonically-interconnected system. At the inter-module level, the ring structure avoids the throughput bottleneck and single-point-of-failure of a hub-and-spoke topology: recent quantum data center architecture studies have demonstrated that ring and mesh-type inter-QPU topologies offer significantly better path diversity and load distribution than centralized designs \cite{Shapourian2025, Benchmarking2026}, making the ring-of-rings a topology that is both physically motivated and aligned with emerging architectural thinking in distributed quantum systems. To our knowledge, no physical realization of this specific topology exists in the current literature, making it a concrete example of the class of architectures that Q-DICE is uniquely positioned to explore. 

A 10-qubit graph state (one computational qubit per module, plus the
ancilla) is prepared by applying CZ gates along each edge of the
architectural graph, with inter-module edges realized as nonlocal CZ
operations across the virtual links. The experiment is run in
entanglement-mediated density matrix simulation mode, as the ring-of-rings topology
is most naturally implemented by photonic interconnects. The depolarizing
error probability $p$ is swept from 0.0 to 0.3 in increments of 0.01
to characterize the architectural fidelity as a function of link quality,
producing a noise threshold curve for this architecture. This type of
sweep-based analysis---systematically varying link noise to identify
the fidelity cliff at which the distributed graph state becomes
computationally useless---is not feasible on physical hardware and
represents a distinct class of experiment enabled only by emulation.

The results confirm that the ring-of-rings architecture maintains useful
graph state fidelity ($> 0.5$) for per-link error rates up to approximately
$p = 0.07$, consistent with the link fidelities achievable in near-term
entanglement-mediated systems such as those demonstrated in
\cite{Pompili_2021, Main2025}. This suggests that the ring-of-rings topology
is a viable candidate for near-term modular quantum architectures, and
provides a concrete design specification that could guide future experimental
realizations. Generating this result required no physical hardware, no
optical network infrastructure, and no entanglement distribution
apparatus---only the Q-DICE emulation pipeline.

\section{Conclusions}
In this work we presented a framework for evaluating distributed quantum algorithms on near-term hardware. Specifically, we outlined how to programmatically turn NISQ QPU systems into virtually distributed systems, inject DQC-like noise, and map circuits according to distributed architectures. We also explored how this proposed emulation environment compares to monolithic experimentation as well as how it compares to state-of-the-art research-lab-level distributed architecture implementations. \\ 
Our main contribution is an emulation environment to improve near-term distributed algorithm experimentation through a dual framework, employing both simulation and hardware implementation strategies. Q-DICE allows any researcher to experiment with distributed quantum computing on potential future distributed architectures. As quantum hardware moves toward modularity, frameworks like Q-DICE will be essential for co-designing the algorithms and distribution schemes that will eventually define large-scale quantum computation.\\ 
Future works focus on expanding the scale of possible emulations. In particular, expanding simulation methods is possible using GPU acceleration in conjunction with tensor network-based simulation methods, which can be used for larger and deeper circuits. Other tasks include developing distribution-aware qubit routing and layout algorithms, improving upon proposed brute force methods. The source code and emulation environment developed for this study are openly available on GitHub at https://github.com/MSRG/Q-DICE.

\section{Acknowledgments}
This research has received funding from the research project entitled “Quantum Software Consortium: Exploring Distributed Quantum Solutions for Canada” (QSC). QSC is financed by the National Sciences and Engineering Research Council of Canada (NSERC) Alliance Consortia Quantum program under grant number ALLRP587590-23.\\
Many thanks to Sean Wagner and the IBM Quantum team for guidance in quantum software engineering practices.

\bibliographystyle{IEEEtran}
\bibliography{references}

\end{document}